\documentclass[11pt,twoside]{article}
\pdfoutput=1
\usepackage{asp2014}

\aspSuppressVolSlug
\resetcounters

\begin{document}

\title{Abstracting the storage and retrieval of image data at the LSST}

\author{Tim~Jenness$^1$,
James~F.~Bosch$^2$,
Pim~Schellart$^2$,
Kian-Tat~Lim$^3$,
Andrei~Salnikov$^3$,
and
Michelle~Gower$^4$
\affil{$^1$Large Synoptic Survey Telescope, Tucson, AZ, USA; \email{tjenness@lsst.org}}
\affil{$^2$Princeton University, Princeton, NJ, USA}
\affil{$^3$SLAC National Accelerator Laboratory, Menlo Park, CA, USA}
\affil{$^4$National Center for Supercomputing Applications, University of Illinois, Urbana-Champaign, IL, USA}
}

\paperauthor{Tim~Jenness}{tjenness@lsst.org}{0000-0001-5982-167X}{LSST}{Data Management}{Tucson}{AZ}{85719}{U.S.A.}
\paperauthor{James~Bosch}{jbosch@astro.princeton.edu}{0000-0003-2759-5764}{Princeton University}{Department of Astrophysical Sciences}{Princeton}{NJ}{08544}{U.S.A.}
\paperauthor{Pim~Schellart}{}{0000-0002-8324-0880}{Princeton University}{Department of Astrophysical Sciences}{Princeton}{NJ}{08544}{U.S.A.}
\paperauthor{Kian-Tat~Lim}{ktl@slac.stanford.edu}{0000-0002-6338-6516}{SLAC}{}{Menlo Park}{CA}{94025}{U.S.A.}
\paperauthor{Andrei~Salnikov}{salnikov@slac.stanford.edu}{0000-0002-3623-0161}{SLAC}{}{Menlo Park}{CA}{94025}{U.S.A.}
\paperauthor{Michelle~Gower}{mgower@illinois.edu}{0000-0001-9513-6987}{University of Illinois}{NCSA}{Urbana-Champaign}{IL}{61801}{U.S.A.}



\begin{abstract}
  Writing generic data processing pipelines requires that the algorithmic code does not ever have to know about data formats of files, or the locations of those files.
  At LSST we have a software system known as ``the Data Butler,'' that abstracts these details from the software developer.
  Scientists can specify the dataset they want in terms they understand, such as filter, observation identifier, date of observation, and instrument name, and the Butler translates that to one or more files which are read and returned to them as a single Python object.
  Conversely, once they have created a new dataset they can give it back to the Butler, with a label describing its new status, and the Butler can write it in whatever format it has been configured to use.
  All configuration is in YAML and supports standard defaults whilst allowing overrides.
\end{abstract}

\section{Introduction}

The Large Synoptic Survey Telescope \citep{2008arXiv0805.2366I}, being built on Cerro Pach\'{o}n in Chile, will be an automated astronomical survey system that will survey approximately $10,000$~deg$^2$ of the sky every few nights in six optical bands.
The associated Data Management System \citep{2017ASPC..512..279J,2018AAS...23136210O} is required to process the data from this telescope and publish it as nightly alerts and as annual data releases.
The LSST science pipelines \citep[see for example][]{2018PASJ...70S...5B,I12-1_adassxxviii} have been designed such that the algorithmic code is insulated from having to know where data comes from and how it has been serialized.
The Butler is the system mediating the storage and retrieval of data, converting Python objects to data files and data files back to Python objects.

\section{Butler Components}

The Butler consists of a high-level Python API, and three core components: Schema, Registry access, and Datastore.
The relationship of these components is shown in Fig.~\ref{fig:butler}.
The Schema defines the data model for relating datasets to each other and is defined consistently for all datasets and instruments.
The Registry classes allow the data model to be queried and are configurable via plugins to allow different backend database systems to be used.
Finally the Datastore deals with the reading and writing of datasets themselves.
Currently there are datastores for a POSIX file system, an in-memory cache, and chained datastores (where writes go to all datastores and reads pull from the first datastore to return it).
For example, the Datastore configuration system allows certain dataset types to be cached in memory only and not written to a file system.
This allows pipeline tasks to be linked together using the Butler as an intermediary without always incurring a write overhead.
The configuration system allows the user to switch to writing intermediate files instead of caching them whilst debugging without changing any code.
To support the Datastores, ``Formatters'' have to be written to serialize and deserialize Python objects to a variety of configurable data formats.

\articlefigure[width=.6\textwidth]{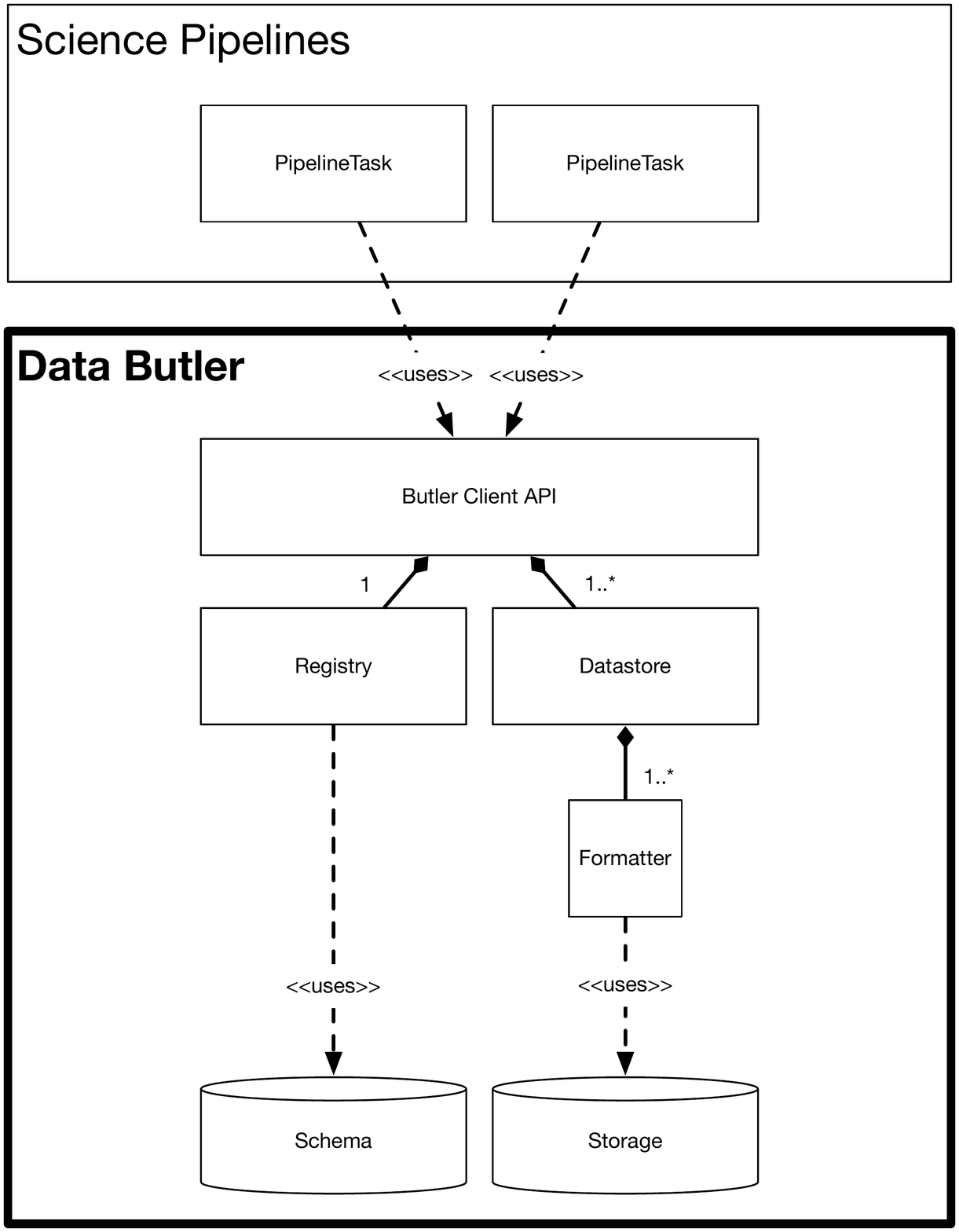}{fig:butler}{Architectural diagram of Butler components.
Datastores use formatters to read from and write to storage, and the registry defines a schema that can be implemented in any database system.
An example of the API is shown in \S\ref{sec:using}.}

\section{Data Model}

The Butler data model is designed to reflect the relationships between observations and calibrations, and also how the sky can be segmented into different regions, associating each dataset with a particular sky region.
This allows you to ask which datasets are needed to calibrate another dataset, which datasets were taken with this filter between these dates, or which datasets would be needed to make a coadd covering this patch of sky.
It can also answer provenance queries, such as asking which coadds in a particular filter had at least this number of observations contributing.
We are designing the Schema to be generally applicable for astronomical data and we are taking into account that we would like to map our schema to ObsCore \citep{2017ivoa.spec.0509L} and CAOM-2 \citep{2012ASPC..461..339D} data models in the future.

\section{Using the Butler}
\label{sec:using}

Individual pipeline tasks work with Python objects.
They put datasets and retrieve datasets from the Datastores.
The Butler maps a Python object  to  a serialization format through a ``StorageClass'' defined in the YAML configuration files for each Datastore.
Changing the serialization format from FITS to HDF5 does not require any code changes for the user and is as simple as editing one line in the configuration file.
Pre-defined components of a dataset, such as the WCS solution, can be retrieved without reading the full dataset if supported by the formatter.
The components supported by each dataset type are defined at the StorageClass level, with code having to be written to assemble a Python composite object from the components and to disassemble a Python object into components.

Below is some user code for retrieving a raw HSC observation along with the relevant flatfield, processing it in some way, and then storing a new version with a different dataset type name.
Calibration datasets can be retrieved by knowing the dataset that is to be calibrated.

\begin{quote}
\begin{small}
\begin{verbatim}
from lsst.daf.butler import Butler

# Configure a new butler
butler = Butler("config.yaml")

# Specify the requested observation via metadata
dataId = {"instrument": "HSC", "obsid": "HSCA04090000"}

# Retrieve the raw data, process it, and store with new label
raw = butler.get("raw", dataId)
flat = butler.get("flat", dataId)
new = doSomething(raw, flat)
butler.put(new, "newlabel", dataId)

# Get just the WCS without reading the full dataset
wcs = butler.get("newlabel.wcs", dataId)
\end{verbatim}
\end{small}
\end{quote}

\section{Header Translation}

To be able to ingest instrument data into a Butler repository, the Butler has to understand some properties of the instrument including filters, detector information, and how to extract metadata from data headers.
We have written a separate Python package, \texttt{astro\_metadata\_translator}, to support header translation and metadata extraction for astronomical instrument headers.
The design of this new package has been influenced by the header translator written for ORAC-DR \citep{2015A&C.....9...40J} and unifies the translation systems previously in use at LSST.
New translators must be written to allow the Butler to understand data during ingest.
Currently, translators exist for DECam, CFHT MegaPrime, and SuprimeCam and Hyper-SuprimeCam from Subaru, with support for LSST test data being added as needed.
This package solely depends on Astropy \citep{2018AJ....156..123A} and does not need any LSST infrastructure.

\begin{quote}
\begin{small}
\begin{verbatim}
from astropy.io import fits
from astro_metadata_translator import ObservationInfo

hdul = fits.open("hsc.fits")
obsInfo = ObservationInfo(hdul[0].header)
print(f"instrument={obsInfo.instrument}, "
      f"date-obs={obsInfo.datetime_begin}")
\end{verbatim}
\end{small}
\end{quote}

\section{Summary}

The Butler frees you from the worry of file formats and file systems when your main concern is processing and characterizing datasets.
The Butler system is not LSST-specific, is written entirely in Python 3 (requiring Python 3.6 or newer following the project baseline \citep{P9-123_adassxxvii}), and is driven by external configuration to suit different use cases.
The Butler, currently undergoing heavy development and considered to be pre-beta, will be released at the end of 2018 alongside v17.0 of the LSST Science Pipelines.
The source code for the Butler can be found at \url{https://github.com/lsst/daf_butler}, and the source code for the header translator can be found at \url{https://github.com/lsst/astro_metadata_translator}.

\acknowledgements We thank Simon Krughoff, Gregory Dubois-Felsmann, and Meredith Rawls for their reviews of the draft paper.
This material is based upon work supported in part by the National Science Foundation through Cooperative Agreement 1258333 managed by the Association of Universities for Research in Astronomy (AURA), and the Department of Energy under Contract No.\ DE-AC02-76SF00515 with the SLAC National Accelerator Laboratory.
Additional LSST funding comes from private donations, grants to universities, and in-kind support from LSSTC Institutional Members.

\bibliography{P13-7}  

\end{document}